\documentclass[a4paper,12pt]{article}
\usepackage[usenames]{color}
\usepackage{amsmath,amssymb}
\usepackage{graphicx}
\numberwithin{equation}{section}
\newcommand{\abs}[1]{\lvert#1\rvert}
\newcommand{\Abs}[1]{\left\lvert#1\right\rvert}
\newcommand{\p}{\partial}

\newcommand{\ti}{\tilde}
\newcommand{\al}{\alpha}
\newcommand{\be}{\beta}

\newcommand{\de}{\delta}
\newcommand{\De}{\Delta}
\newcommand{\ep}{\epsilon}
\newcommand{\ka}{\kappa}
\newcommand{\si}{\sigma}
\newcommand{\la}{\lambda}
\newcommand{\La}{\Lambda}
\newcommand{\om}{\omega}
\newcommand{\sgn}{\mathop{\mathrm{sgn}}\nolimits}

\newcommand{\calO}{\mathcal{O}}
\newcommand{\bra}[1]{\langle#1|}
\newcommand{\ket}[1]{|#1\rangle}
\newcommand{\bracket}[2]{\langle#1|#2\rangle}
\newcommand{\ev}[1]{\langle #1 \rangle}
\newcommand{\Ev}[1]{\left\langle #1 \right\rangle}
\newcommand{\dket}[1]{|#1\rangle\!\rangle}
\newcommand{\nt}{\notag\\}
\newcommand{\we}{\wedge}

\newcommand{\cP}{\mathcal{P}}
\newcommand{\normalorder}[1]{{}_\circ^\circ#1{}_\circ^\circ}
\newcommand{\m}{\mathrm{m}}
\newcommand{\Sym}{\mathop{\mathrm{Sym}}\nolimits}
\newcommand{\bssym}[1]{\bs\Sym\bigl[\emV;#1\bigr]}
\newcommand{\mV}{2i\ka[b_0^-,O]}
\newcommand{\emV}{e^{\mV}}
\newcommand{\bs}{\bra{B}}
\newcommand{\bsv}{\bra{B}\emV}
\newcommand{\bsla}{\bra{B;\la}}
\newcommand{\Bs}{\bra{B^-}}
\newcommand{\Bsv}{\bra{B^-}\emV}
\newcommand{\Bsla}{\bra{B^-;\la}}
\newcommand{\Bssym}[1]{\Bs\Sym\bigl[\emV;#1\bigr]}
\newcommand{\VLPP}{{V_\text{LPP}^{123}}}
\newcommand{\LPP}[3]{\ev{\VLPP\,|#1}_1\ket{#2}_2\ket{#3}_3}
\newcommand{\LPPT}[3]{\ev{\VLPP;T|#1}_1\ket{#2}_2\ket{#3}_3}
\newcommand{\sfr}{S_c^{\text{free}}}
\newcommand{\sbsft}{S_D^{\text{BSFT}}}
\newcommand{\sint}{S_D^{\text{int}}}
\newcommand{\scint}{S_c^{\text{int}}}
\newcommand{\soc}{S_{\text{oc}}}
\newcommand{\delc}{\de_{\La_c}}
\newcommand{\delcm}{\de_{\La_c^-}}
\newcommand{\delo}{\de_{\La_o}}

\renewcommand{\Re}{\mathop{\mathrm{Re}}\nolimits}

\begin{document}

\begin{titlepage}
\title{\vspace{1cm}
On open-closed extension of boundary string field theory}
\author{\vspace{2mm}
Akira Ishida$^1$\thanks{E-mail: \ttfamily ishida.ak@gmail.com}~~and
Shunsuke Teraguchi$^2$\thanks{E-mail: \ttfamily teraguch@ifrec.osaka-u.ac.jp}\\[4mm]
$^1$\textit{\normalsize Department of Physics, Daido University,}\\
\textit{\normalsize 10-3 Takiharu-cho, Minami-ku, Nagoya 457-8530, Japan}\\[3mm]
$^2$\textit{\normalsize WPI Immunology Frontier Research Center,
Osaka University,}\\
\textit{\normalsize Osaka 565-0871, Japan}
}
\date{}
\maketitle
\thispagestyle{empty}
\vspace{5mm}
\begin{abstract}
We investigate a classical open-closed string field theory whose open string
sector is given by boundary string field theory. The open-closed interaction
is introduced by the overlap of a boundary state with a closed
string field.
With the help of the Batalin-Vilkovisky formalism, the closed string sector
is determined to be the HIKKO closed string field theory.
We also discuss the gauge invariance of this theory in both open
and closed string sides.
\end{abstract}

\end{titlepage}

\section{Introduction}
Off-shell formulations of string theory are indispensable for understanding
its nonperturbative aspects.
String field theory (SFT) is a promising off-shell formulation and
recent remarkable progress in open string field theory enables us
to investigate the dynamics of D-branes via tachyon condensation
(for a review, see \cite{Sen:2004nf}).
Most of these studies consider open string field theory on a D-brane.
However, as D-branes are sources of closed strings,
such D-brane systems should be coupled to a closed SFT to give a complete
description of tachyon condensation.

There are several attempts to consider D-branes
in the framework of closed SFT. One way in this direction is
to construct an open-closed SFT by directly introducing both open
and closed string fields \cite{Zwiebach:1997fe,Kugo:1997rm}.
Another approach is to introduce a D-brane boundary state $\bs$ as a source
in a closed SFT. The action of such a system would be given by adding
a source term to a closed SFT action $S_c$:
\begin{equation}
S_c+\bs\,e^{-\int\! d\si V(\si)}(c_0-\ti c_0)\ket\Psi\,,\label{addsource}
\end{equation}
where $\ket\Psi$ is a closed string field. Here
$V(\si)$ is a marginal (on-shell) operator describing open string excitations
on the D-brane and $\si$ parametrizes the boundary of the worldsheet.
In \cite{Hashimoto:1997vz}, the authors considered the HIKKO closed SFT
\cite{Hata:1986kj} with the source term in the presence of a constant
electro-magnetic flux.
They argued that the boundary state should transform under the closed string
gauge transformation and found that the gauge invariance requires to add
the Dirac-Born-Infeld (DBI) action to \eqref{addsource}.
The authors in \cite{Asakawa:2003jv} also considered a closed SFT with
a source term representing a marginal deformation, namely \eqref{addsource}.
They pointed out that an (interacting) closed SFT requires a modification
of the deformed boundary state; marginal deformations, which are expected
to be solutions in open string theory, are no longer
consistent solutions in open-closed string theory.

In most of these previous studies based on boundary states and closed SFT,
the open string excitations have been taken to be on-shell. The aim of this
paper is to generalize those open string fluctuations into off-shell cases.
For this purpose, we adopt boundary string field theory (BSFT)
\cite{Witten:1992qy,Witten:1992cr,Shatashvili:1993ps,BSFT2}.
BSFT, which is a version of open string field theory, is formulated on
the space of all boundary operators including off-shell ones.
Although BSFT is originally defined in terms of a worldsheet sigma model
with boundary interactions, it can be reformulated by using boundary
states \cite{Teraguchi:2006tb,Ishida:2008tj}.
This would suggest that BSFT is more suitable for introducing couplings with
closed SFTs than cubic open SFT \cite{Witten:1985cc}.
We thus seek for an open-closed SFT whose open string sector is BSFT.
We expect that this open-closed SFT is a natural
extension of \cite{Hashimoto:1997vz,Asakawa:2003jv}.
We refer to this open-closed SFT as boundary open-closed string field theory.

The organization of this paper is as follows. In the next section,
we give a short review on the Batalin-Vilkovisky (BV) formalism for
classical open-closed SFT and derive a gauge transformation law.
In the construction of a SFT, a guiding principle is the stringy gauge
invariance.
The BV formalism \cite{BV} is useful for this purpose
since an action satisfying the BV master equation automatically
has gauge invariance. It was proposed in \cite{Kajiura:2005sn}
that a classical open-closed SFT should satisfy the BV classical open-closed
master equations, which is the classical part of the full BV master equation.
We examine how those master equations determine the gauge transformation.
We also provide basic ingredients for the construction of
our open-closed SFT.

In section \ref{oc-interaction}, we couple BSFT to a free closed SFT with
the conventional open-closed interaction as in \eqref{addsource}.
We then add appropriate terms required by the BV classical open-closed
master equations.
In this manner, we find that the joining-splitting type
3-closed-string interaction is required from the BV master equation.
An essential point is the utilization of the factorization property of
conformal field theory (CFT).

However, even with the joining-splitting type vertex, there also exists a
region of string length parameters where the master equation is not satisfied.
For this reason, it seems difficult to find an open-closed SFT action
which completely satisfies the master equation in our framework.
Though this is somewhat disappointing, as demonstrated in section
\ref{gauge}, we can still show that the resulting action has gauge
invariance if we restrict the string length of the gauge parameter to negative.
In the final section, we shall briefly discuss the relation between
our open-closed SFT and the D-brane soliton state in a closed SFT
\cite{Baba:2006rs,Baba:2007tc}.
Our conventions and some useful formulas are summarized in appendices.

\section{Batalin-Vilkovisky formalism and string field theory}\label{BVSFT}
In this section, we briefly review the BV formalism for
classical open-closed SFT
and derive the gauge transformation law for string fields. The term `classical'
represents a theory without loops, namely consisting only of diagrams with
genus zero and at most one boundary.
We also review BSFT, which is our starting point in constructing
our open-closed SFT.
\subsection{Batalin-Vilkovisky formalism and gauge transformation}
In an open-closed SFT,
a closed string field $\Psi$ and an open string field $O$
are expanded in terms of the bases $\{\Psi_I\}$
and $\{O_i\}$ respectively,
\begin{equation}
\Psi=\sum_I\Psi_I\psi^I,\qquad
O=\sum_iO_i\la^i\,,
\end{equation}
where $\psi^I$ and $\la^i$ are target space closed and open string fields,
respectively. Throughout this paper, we denote open and
closed string fields by the indices $i,j,\cdots$ and $I,J,\cdots$,
respectively. A key ingredient in the BV formalism
is a fermionic two-form $\om=\om_c+\om_o$,
\begin{equation}
\om_c=-d\psi^I\we\om_{IJ}^cd\psi^J\,,\qquad
\om_o=-d\la^i\we\om_{ij}^od\la^j\,,
\end{equation}
which is non-degenerate and closed.
The anti-bracket is defined as the sum of those in the open and closed
string sectors
\begin{equation}
\{A,B\}=\{A,B\}_c+\{A,B\}_o\,,
\end{equation}
where
\begin{equation}
\{A,B\}_c=\frac{\p_rA}{\p\psi^I}\om_c^{IJ}\frac{\p_lB}{\p\psi^J}\,,\qquad
\{A,B\}_o=\frac{\p_rA}{\p\la^i}\om_o^{ij}\frac{\p_lB}{\p\la^j}\,.
\end{equation}
Here $\frac{\p_l}{\p\psi}$ and $\frac{\p_r}{\p\psi}$ are
the left and right derivatives respectively, and
$\om_c^{IJ}$ and $\om_o^{ij}$ are the inverse matrices of
$\om^c_{IJ}$ and $\om^o_{ij}$, respectively.

In the BV formalism,
a classical closed SFT action $S_c(\psi)$ obeying the classical master
equation,\footnote{The classical master equation can be understood as a
classical truncation of the quantum master equation, $\frac12\{S,S\}+\De S=0$.
In the construction of classical SFT, we can ignore $\De=\De_c+\De_o$ because
$\De_c$ increases the genus $g$ and $\De_o$
increases the number of boundaries $b$ for $b>0$.
We should note that, though the BV master equation is a strong
consistency condition for SFTs, it does not necessarily guarantee
a single cover of the moduli space.}
\begin{equation}
\{S_c,S_c\}_c=0\,,\label{BVc}
\end{equation}
is automatically guaranteed to possess a gauge invariance;
the action is invariant under the closed string gauge transformation
\begin{equation}
\de_c\psi^I=\left(
\om_c^{IJ}\frac{\p_r}{\p \psi^K}\frac{\p_lS_c}{\p \psi^J}
+\frac12(-1)^{JK}\frac{\p_r\om_c^{IJ}}{\p\psi^K}
\frac{\p_lS_c}{\p\psi^J}\right)\ep^K,\label{closedgaugetr}
\end{equation}
as seen from the derivative of the master equation \eqref{BVc}:
\begin{equation}
0=\frac12\frac{\p_r\{S_c,S_c\}_c}{\p\psi^K}\ep^K
=\frac{\p_rS_c}{\p\psi^I}\left(\om_c^{IJ}\frac{\p_r}{\p\psi^K}
\frac{\p_lS_c}{\p\psi^J}+\frac12(-1)^{JK}\frac{\p_r\om_c^{IJ}}{\p\psi^K}
\frac{\p_lS_c}{\p\psi^J}\right)\ep^K.
\end{equation}
Here, $\ep$ is an infinitesimal parameter and $(-1)^I$ is the Grassmann
parity of $\psi^I$.
Similarly, an open SFT action $S_o(\la)$ can be constructed to be
a solution of the master equation
\begin{equation}
\{S_o,S_o\}_o=0\,.\label{BVo}
\end{equation}
The same argument as above leads to the gauge transformation law
in the open string side
\begin{equation}
\de_o\la^i=\left(
\om_o^{ij}\frac{\p_r}{\p \la^k}\frac{\p_lS_o}{\p\la^j}
+\frac12(-1)^{jk}\frac{\p_r\om_o^{ij}}{\p\la^k}
\frac{\p_lS_o}{\p\la^j}\right)\ep^k\,.\label{gauge_OSFT}
\end{equation}

However, once open-closed interactions are taken into account,
the master equations above are not a consistent classical truncation.
To see this, let us consider a classical open-closed SFT action of the form
\begin{equation}
\soc(\psi,\la)=S_c(\psi)+S_D(\psi,\la)\,,
\end{equation}
where $S_c$ is a classical closed SFT action and $S_D$ corresponds to disks
with and without closed string insertions. The anti-bracket of $\soc$
with itself is given by
\begin{equation}\label{socsoc}
\{\soc,\soc\}=\{S_c,S_c\}_c+2\{S_c,S_D\}_c+\{S_D,S_D\}_o+\{S_D,S_D\}_c\,.
\end{equation}
The last term represents a surface with two boundaries
and should be ignored in classical SFT.
Therefore, as proposed in \cite{Kajiura:2005sn},
the BV classical open-closed master equations are given by
the following two equations:\footnote{The authors of
\cite{Kajiura:2005sn}
investigated the classical part of
Zwiebach's open-closed SFT \cite{Zwiebach:1997fe} in terms of homotopy algebra
and defined open-closed homotopy algebra (OCHA).
A quantum generalization of OCHA was recently discussed in
\cite{Munster:2011ij}.}
\begin{align}
0=&\{S_c,S_c\}_c\,,&(b=0) \label{OCBV0}\\
0=&2\{S_c,S_D\}_c+\{S_D,S_D\}_o\,,&(b=1) \label{OCBV1}
\end{align}
where we have split \eqref{socsoc} into two equations for
each number of boundaries $b$.

Similarly to the case above, we can derive gauge transformations by applying
$\frac{\overleftarrow\p}{\p\psi^K}\ep^K+\frac{\overleftarrow\p}{\p\la^k}\ep^k$
on \eqref{OCBV0} and \eqref{OCBV1}.
We see from \eqref{OCBV0} that the gauge transformation of $S_c$ is the
same as in the closed SFT. This also requires that the closed string field
in $S_D$ should transform as in \eqref{closedgaugetr}. For \eqref{OCBV1},
we have
\begin{align}
0&=\frac{\p_r}{\p\psi^K}\left[\{S_c,S_D\}_c+\frac12\{S_D,S_D\}_o\right]\ep^K
+\frac{\p_r}{\p\la^k}\left[\{S_c,S_D\}_c+\frac12\{S_D,S_D\}_o\right]\ep^k\nt
&=\frac{\p_rS_D}{\p\la^i}\left\{
\left[\om_o^{ij}\frac{\p_r}{\p\la^k}\frac{\p_lS_D}{\p\la^j}
+\frac12(-1)^{jk}\frac{\p_r\om_o^{ij}}{\p\la^k}\frac{\p_lS_D}{\p\la^j}
\right]\ep^k
+\om_o^{ij}\left(\frac{\p_r}{\p\psi^K}
\frac{\p_lS_D}{\p\la^j}\right)\ep^K\right\}\nt
&\quad+\frac{\p_rS_D}{\p\psi^I}\left[
\om_c^{IJ}\frac{\p_r}{\p\psi^K}\frac{\p_lS_c}{\p\psi^J}
+\frac12(-1)^{JK}\frac{\p_r\om_c^{IJ}}{\p\psi^K}
\frac{\p_lS_c}{\p\psi^J}\right]\ep^K\nt
&\quad+\frac{\p_rS_c}{\p\psi^I}
\left\{\left[\om_c^{IJ}\frac{\p_r}{\p\psi^K}\frac{\p_lS_D}{\p\psi^J}
+\frac12(-1)^{JK}\frac{\p_r\om_c^{IJ}}{\p\psi^K}
\frac{\p_lS_D}{\p\psi^J}\right]\ep^K
+\om_c^{IJ}\left(\frac{\p_r}{\p\la^k}\frac{\p_lS_D}{\p\psi^J}\right)\ep^k
\right\}.\label{OCBV11}
\end{align}
Here we have collected terms proportional to $(\p_rS_D/\p\la^i)$
in the second line, which determine the gauge transformation for
open string fields. The third line is formed
to obtain the consistent gauge transformation with $S_c$.
The last line is the collection of the remaining terms, which
imposes a constraint on the infinitesimal parameters
\begin{equation}\label{constraint}
\left[\om_c^{IJ}\frac{\p_r}{\p\psi^K}\frac{\p_lS_D}{\p\psi^J}
+\frac12(-1)^{JK}\frac{\p_r\om_c^{IJ}}{\p\psi^K}
\frac{\p_lS_D}{\p\psi^J}\right]\ep^K
+\om_c^{IJ}\left(\frac{\p_r}{\p\la^k}\frac{\p_lS_D}{\p\psi^J}\right)\ep^k=0\,.
\end{equation}
We therefore find that the total action $S_c+S_D$ is invariant under
the following gauge transformations,
\begin{align}
\de_c\psi^I&=\left[\om_c^{IJ}\frac{\p_r}{\p\psi^K}\frac{\p_lS_c}{\p\psi^J}
+\frac12(-1)^{JK}\frac{\p_r\om_c^{IJ}}{\p\psi^K}
\frac{\p_lS_c}{\p\psi^J}\right]\ep^K\,,\quad
\de_c\la^i
=\om_o^{ij}\left(\frac{\p_r}{\p\psi^K}\frac{\p_lS_D}{\p\la^j}\right)\ep^K\,,
\label{closed}\\
\de_o\la^i&=\left[\om_o^{ij}\frac{\p_r}{\p\la^k}\frac{\p_lS_D}{\p\la^j}
+\frac12(-1)^{jk}\frac{\p_r\om_o^{ij}}{\p\la^k}
\frac{\p_lS_D}{\p\la^j}\right]\ep^k\,,\label{open}
\end{align}
where the infinitesimal parameters must be chosen to satisfy
\eqref{constraint}.
The second equation in \eqref{closed} shows that open string fields
get transformed by the closed string gauge transformation.
We apply this formalism to construct a classical open-closed SFT
whose open string part is BSFT.
\subsection{Closed and open string field theories}
We give some details of closed and open string fields in our construction.
To begin with, it is convenient to define the following ghost zero modes
\begin{equation}
b_0^-=\frac12(b_0-\ti b_0)\,,\qquad c_0^-=c_0-\ti c_0\,,
\end{equation}
which satisfy the anti-commutation relation\footnote{Throughout this paper,
the bracket denotes the graded commutator $[A,B]\equiv AB-(-1)^{AB}BA$.}
\begin{equation}
[b_0^-,c_0^-]=1\,.
\end{equation}
We adopt closed string fields as
a conventional one satisfying the conditions,
\begin{equation}
b_0^-\ket\Psi=0\,,\qquad L_0^-\ket\Psi=0\,,\label{b0L0}
\end{equation}
where $L_0^-\equiv L_0-\ti L_0$. The second condition imposes
the level matching condition. This condition can also be represented as
$\cP\ket\Psi=\ket\Psi$, where the projector $\cP$ is defined by
\begin{equation}\label{Proj}
\cP=\int_0^{2\pi}\frac{d\theta}{2\pi}e^{i\theta(L_0-\ti L_0)}\,.
\end{equation}
On the other hand, we take open string fields as boundary operators.
The ghost number of the closed string field $\ket\Psi$ is 2,
while that of the open string field $O$ is 1 as usual.
They are expanded in terms of target space string fields as
\begin{equation}
\ket\Psi=\sum_I\ket{\Psi_I}\psi^I,\qquad
O=\sum_iO_i\la^i,
\end{equation}
where $\ket{\Psi_I}$ and $O_i$ are bases of string fields of
any worldsheet ghost number.
The statistics of target space fields are taken to satisfy the
correct statistics of string fields, namely,
\begin{equation}
(-1)^I\equiv\Abs{\psi^I}=\Abs{\ket{\Psi_I}},\qquad
(-1)^i\equiv\abs{\la^i}=-\abs{O_i}\,.
\end{equation}
Similarly, we assign a spacetime ghost number $g$ to target space fields
so that string fields have the correct ghost number:
\begin{equation}
g(\psi^I)=2-G(\Psi_I),\qquad g(\la^i)=1-G(O_i)\,,
\end{equation}
where $G$ represents the worldsheet ghost number.

By using the ingredients above, we define
the closed 2-form for closed strings as
\begin{equation}
\om_c=\bra{d\Psi}c_0^-\ket{d\Psi}\,,
\end{equation}
where $\ket{d\Psi}=\ket{\Psi_I}d\psi^I$, and $\bra{d\Psi}$ denotes
its BPZ conjugate.
Thus the kinetic term for closed string fields is given by the canonical one:
\begin{equation}
\sfr=\frac12\bra{\Psi}c_0^-Q_B\ket{\Psi}\,.\label{Sfree}
\end{equation}

In the rest of this section, we focus on BSFT.
We mainly follow the formulation in \cite{Teraguchi:2006tb,Ishida:2008tj},
where BSFT is rewritten in terms of
boundary state, since the boundary state formalism is a suitable framework
to consider the coupling between open and closed strings.
The fundamental object in BSFT is the operator $O$ representing
a boundary operator $\calO(\si)$ after $\si$ integration,
\begin{equation}
O\equiv\int_{-\pi}^{\pi}\frac{d\si}{2\pi}\calO(\si)\,,
\end{equation}
Since $L_0^-$ generates a rotation of the disk, the boundary operator $O$
satisfies
\begin{equation}
[L_0^-,O]=0\,.
\end{equation}
With the off-shell boundary operator $O$, deformed boundary states
\begin{equation}
\bsla\equiv\bsv\,,
\end{equation}
play an important role. Here $\bs$ represents a D-brane boundary state
and $\ka$ is the open string coupling constant.
Although conformal symmetries are lost
due to the off-shell deformation, $\bsla$ is still annihilated by
$b_0^-$ and $L_0^-$,
\begin{equation}
\bsla b_0^-=0\,,\qquad \bsla L_0^-=0\,,
\end{equation}
which takes the same form as the condition for closed string fields in
\eqref{b0L0}.
The closed two-form $\om_o$ is defined by two-point correlation functions
of the deformed worldsheet theory, which is given by
\begin{equation}
\om_o=\bssym{dO,dO}\ket0\,,
\end{equation}
where $dO=O_i d\la^i$ and $\ket0$ is the $SL(2,\mathbb{C})$ vacuum.
The symbol $\Sym[\cdots]$ is defined by
\begin{align}
&\Sym[e^{-V};O_1,O_2,\cdots,O_n]\nt
&\quad=\int_0^1dt_1 \int_{t_1}^1dt_2 \cdots \int_{t_{n-1}}^1 dt_n
e^{-t_1V}O_1e^{-(t_2-t_1)V}O_2\cdots O_n e^{-(1-t_n)V}\pm(\text{perms})\,.
\end{align}
In the construction of BSFT, we need a fermionic vector $V^i$.
As shown in \cite{Witten:1992qy}, if the vector $V$ is nilpotent
and generates a symmetry of $\om_o$, i.e.,
\begin{equation}
(di_V+i_Vd)\om_o=d(i_V\om_o)=0\,,
\end{equation}
the BSFT action defined by $d\sbsft=i_V\om_o$ automatically satisfies the
master equation \eqref{BVo}. It was shown that
the vector generated by the bulk BRST operator $Q_B$,
\begin{equation}
\de_V O\equiv[Q_B,O]\,,\label{defofV}
\end{equation}
indeed satisfies the requirement above. Note that since
$\de_V O=\frac{\p_rO}{\p\la^i}V^i=O_iV^i$, the component form of \eqref{defofV}
is
\begin{equation}
O_iV^i=[Q_B,O]\,,\label{compV}
\end{equation}
which we use to calculate anti-brackets in the subsequent sections.
By using the vector $V$, the BSFT action is given by
\begin{equation}
d\sbsft=-\bssym{dO,[Q_B,O]}\ket0\,.\label{defofS}
\end{equation}
The integration of $dS$ was performed in
\cite{Teraguchi:2006tb,Shatashvili:1993ps}, and the result is
\begin{equation}
\sbsft=-\frac1{4\ka^2}\bsv c_0^-Q_Bc_0^-\ket0
+\frac{i}{2\ka}\bssym{[Q_B,O]}c_0^-\ket0\,.\label{act-BSFT}
\end{equation}
The gauge transformation can be read off from \eqref{gauge_OSFT}
and is given by \cite{Ishida:2008tj}
\begin{equation}
\de_{\La} O=[Q_B,\Lambda]
+i\ka\bssym{[Q_B,O],[b_0^-,\Lambda],O_i}\ket{0}\,\omega_o^{ij}O_j\,,
\label{gaugeBSFT}
\end{equation}
where $\La$ is a gauge parameter of ghost number 0. As we shall see later,
once we take open-closed interactions into account, $\La$ is no longer an
arbitrary parameter and the constraint on $\La$ eliminates the second term of
\eqref{gaugeBSFT}.
\section{Coupling BSFT to closed SFT}
\label{oc-interaction}
We construct an open-closed SFT action by requiring the action to
satisfy the BV classical open-closed master equations.
We begin with the BSFT action and the free closed SFT action
as well as the conventional open-closed interaction term given by
the overlap of an off-shell boundary state with a closed string field.
We shall see that the joining-splitting type 3-closed-string
interaction is necessary.
\subsection{Our ansatz and BV master equations}
We start with the ansatz
\begin{equation}
\sfr+\sint+\sbsft\,,
\end{equation}
and subsequently add appropriate terms required by the master equations.
Here $\sfr$ is the kinetic term for closed string fields given in
\eqref{Sfree} and $\sint$ describes the interaction between
closed strings and boundary states:\footnote{This interaction term
between open and closed string fields is analogous to the ones considered
in \cite{Moeller:2010mh}.
The authors introduced the interaction terms into cubic open SFT
to investigate closed string cohomology.}
\begin{equation}
\sint=\frac12\bsv c_0^-\ket\Psi\,.\label{Sint}
\end{equation}

Let us check to what extent the ansatz satisfies
the BV classical open-closed master equations
\eqref{OCBV0} and \eqref{OCBV1}.
The two-form $\om$ has already been defined in the previous section,
whose components are
\begin{equation}
\om_{IJ}^c=(-1)^{I+1}\bra{\Psi_I}c_0^-\ket{\Psi_J}\,,\qquad
\om_{ij}^o=(-1)^{i+1}\bssym{O_i,O_j}\ket0\,.
\end{equation}
Some useful formulas relevant for the calculation below are summarized
in Appendix \ref{convention}.
As a warm-up, let us show that the kinetic term satisfies the first equation
\eqref{OCBV0}.
Using the completeness condition \eqref{complete}, we have
\begin{align}
\{\sfr,\sfr\}_c
&=\bra{\Psi}c_0^-Q_B\ket{\Psi_I}\om_c^{IJ}(-1)^J
\bra{\Psi_J}c_0^-Q_B\ket{\Psi}\nt
&=-\bra{\Psi}c_0^-Q_Bb_0^-c_0^-Q_B\ket{\Psi}\nt
&=-\bra{\Psi}c_0^-Q_B^2\ket{\Psi}
+\frac12\bra{\Psi}c_0^-Q_Bc_0^-L_0^-\ket{\Psi}
-\bra{\Psi}c_0^-Q_Bc_0^-Q_Bb_0^-\ket{\Psi}\nt
&=0\,,
\end{align}
where we moved the $b_0^-$ to the right.
For the second equation \eqref{OCBV1}, we have
\begin{align}
2&\{\sbsft+\sint,\,\sfr\}_c
+\{\sbsft+\sint,\,\sbsft+\sint\}_o\nt
&=\{\sbsft,\sbsft\}_o&(n=0)\nt
&\quad+2\{\sint,\sfr\}_c+2\{\sint,\sbsft\}_o&(n=1)\nt
&\quad+\{\sint,\sint\}_o\,,&(n=2)
\end{align}
where $n$ denotes the number of closed string insertions in the
corresponding term.
The term without closed strings vanishes by construction.
The terms with $n=1$ are evaluated as
\begin{align}
&2\{\sint,\sfr\}_c+2\{\sint,\sbsft\}_o\nt
&=\bsv c_0^-\ket{\Psi_I}\om_c^{IJ}(-1)^J\bra{\Psi_J} c_0^-Q_B\ket\Psi
+2i\ka\bssym{[O_i,b_0^-]}c_0^-\ket\Psi\om_o^{ij}\om^o_{jk}V^k\nt
&=-\bsv c_0^-b_0^-c_0^-Q_B\ket\Psi
+2i\ka\bssym{[O_i,b_0^-]}c_0^-\ket\Psi V^i\nt
&=-\bsv c_0^-Q_B\ket\Psi+2i\ka\bssym{[[O,Q_B],b_0^-]}c_0^-\ket\Psi\nt
&=-\bsv c_0^-Q_Bb_0^-c_0^-\ket\Psi
-2i\ka\bssym{[[b_0^-,O],Q_B]}c_0^-\ket\Psi\nt
&=\bsv Q_Bc_0^-\ket\Psi-\bsv Q_Bc_0^-\ket\Psi=0\,,\label{B1n1}
\end{align}
where we used \eqref{delSBSFT} in the second line and inserted
$1=[b_0^-,c_0^-]$ in the second last line.
It is worth noting the implication of the above equation \eqref{B1n1}.
If one constructs open and closed string field theories with respect to
different BRST operators, the BRST operator in the first term comes from
the closed string side, while the second one from the open string side.
Therefore \eqref{B1n1} is regarded as a compatibility condition for
the two BRST operators.

The remaining one is the term with $n=2$,
\begin{align}
\{\sint,\sint\}_o&=\ka^2\bssym{O_i}\ket\Psi\,\om_o^{ij}
(-1)^{j+1}\bssym{O_j}\ket\Psi\,,\label{b1n2}
\end{align}
which does not vanish.
This implies an additional closed string interaction of the form
\begin{align}
\scint=\frac{\ka^2}3\bra\Psi c_0^-\ket{\Psi\star\Psi}\,.\label{scint}
\end{align}
Here we do not specify the detail of the star product yet.
We only assume the cyclic property
\begin{equation}
\bra Ac_0^-\ket{B\star C}=(-1)^{(A+B)C}\bra Cc_0^-\ket{A\star B}
=(-1)^{(B+C)A}\bra Bc_0^-\ket{C\star A}\,,\label{cyclic}
\end{equation}
and that the product of two string fields is a string field satisfying
\eqref{b0L0}.
The contribution coming from the interaction term \eqref{scint}
to the master equation is
\begin{align}
2\{\sint,\scint\}_c
&=\ka^2\bsv c_0^-\ket{\Psi_I}\om_c^{IJ}(-1)^J
\bra{\Psi_J} c_0^-\ket{\Psi\star\Psi}\nt
&=-\ka^2\bsv c_0^-b_0^- c_0^-\ket{\Psi\star\Psi}\nt
&=-\ka^2\bsv c_0^-\ket{\Psi\star\Psi}\,,
\end{align}
and we require this to cancel \eqref{b1n2}:
\begin{equation}
0=(-1)^{j+1}\bssym{O_i}\ket\Psi\om_o^{ij}
\bssym{O_j}\ket\Psi -\bsv c_0^-\ket{\Psi\star\Psi}\,.\label{BV12}
\end{equation}
Thus we found that three-closed-string interaction is necessary.
This is consistent with \cite{Hashimoto:1997vz},
where the gauge transformation of open string fields is written
in terms of the star product of the HIKKO closed SFT.
We shall look for an appropriate star product,
namely a three-string interaction vertex. We should also keep in mind
that the resulting closed SFT action must satisfy the master equation
\eqref{OCBV0}.
\subsection{Three-closed-string vertex}\label{3vertex}
We have seen that the master equation requires
a specific form of 3-closed-string vertex as in \eqref{BV12}.
In this subsection, we discuss this issue further.
The first term of \eqref{BV12} is reminiscent of the factorization property
in CFT. Suppose a surface with a thin neck. When it is pinched and
the surface is separating into two surfaces, a correlation function
factors into the product of those on two surfaces.
We shall see that such a factorization indeed occurs in the second term
in \eqref{BV12} if we adopt the HIKKO star product.

\begin{figure}[ht]
\centering
\includegraphics{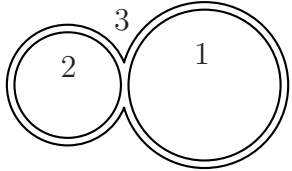}
\begin{picture}(0,0)
\put(-40,40){1}
\put(-90,35){2}
\put(-70,53){3}
\end{picture}
\caption{Three closed string interaction in the HIKKO closed SFT for
$\abs{\al_3}=\abs{\al_1}+\abs{\al_2}$.
The parameters $\al_i$ $(i=1,2,3)$ denote the lengths of each string.}
\label{figvertex}
\end{figure}

In the HIKKO closed SFT \cite{Hata:1986kj}, we need to introduce the string length
parameter $\al$. We write it explicitly as
\begin{equation}
\ket{\Psi(\al)}\equiv\ket\Psi\otimes\ket\al\,,
\end{equation}
where the normalization is $\ev{\al_1|\al_2}=2\pi\de(\al_1-\al_2)$.\footnote{
Note that the BPZ conjugate of $\ket\al$ is $\bra{-\al}$. }
The HIKKO 3-closed-string vertex is defined by the overlap of three closed
strings as depicted in Figure \ref{figvertex}.

In the following, we use the three-string vertex defined in CFT language, which
is called the LPP vertex \cite{LPP}. Since the total central charge vanishes,
we implicitly use the generalized
gluing and resmoothing theorem \cite{LPP2,Asakawa:1998dv}.
The star product is determined through
the LPP vertex $\bra\VLPP$ as follows:
\begin{equation}\label{defstar}
\bra{\Psi_1(\al_1)}c_0^-\ket{\Psi_2(\al_2)*\Psi_3(\al_3)}
=2\pi\de(\al_1+\al_2+\al_3)\LPP{\Psi_1(\al_1)}{\Psi_2(\al_2)}{\Psi_3(\al_3)}\,.
\end{equation}
In this prescription, the LPP vertex is defined in terms of conformal mappings
$h_r$ $(r=1,2,3)$ from three unit disks (with coordinates $w_r$)
to a complex plane (with a coordinate $z$):
\begin{equation}
\LPP{\Psi_1(\al_1)}{\Psi_2(\al_2)}{\Psi_3(\al_3)}
=\Ev{h_1[\Psi_1(0)]\,h_2[\Psi_2(0)]\,h_3[\Psi_3(0)]}\,,\label{defLPP}
\end{equation}
where the operators $\Psi_r(w_r=0)$ are defined through the relation
$\ket{\Psi_r}_r=\Psi_r(0)\ket0_r\,$. Here $h_r[\Psi_r(0)]$ is the conformal
transformation of $\Psi_r(0)$ by $h_r$. For instance, if $\phi$ is a primary
of dimension $d$, $f[\phi(0)]=(f'(0))^d\phi(f(0))$.
The basic properties of the HIKKO closed SFT including the explicit
forms of $h_r$ are summarized in Appendix \ref{HIKKO}.

Let us evaluate the second term in \eqref{BV12} by using the LPP vertex
above:
\begin{equation}
\LPP{\Psi(\al_1)}{\Psi(\al_2)}{B(\al_3);\la}\,,\label{PPB}
\end{equation}
where we have assigned the length parameter to the boundary state as well,
\begin{equation}
\ket{B(\al);\la}\equiv\ket{B;\la}\otimes\ket{\al}\,.
\end{equation}
In the following, we focus on the parameter region with $\al_1,\al_2<0$ and
$\al_3=-\al_1-\al_2>0$. In this setting, the strings ``1'' and ``2'' in
Figure \ref{figvertex} represent the two closed strings, while ``3''
corresponds to the boundary state.
Thus the two closed string worldsheets attach to each other
only at a point. In the discussion below, it is important to note that
the whole worldsheet can be considered as splitting into two surfaces.
In order to evaluate \eqref{PPB},
we follow the procedure given in \cite{Baba:2007tc}.
We represent the boundary state as the boundary of a worldsheet
with an appropriate boundary deformation.

\begin{figure}[t]
\centering
\includegraphics{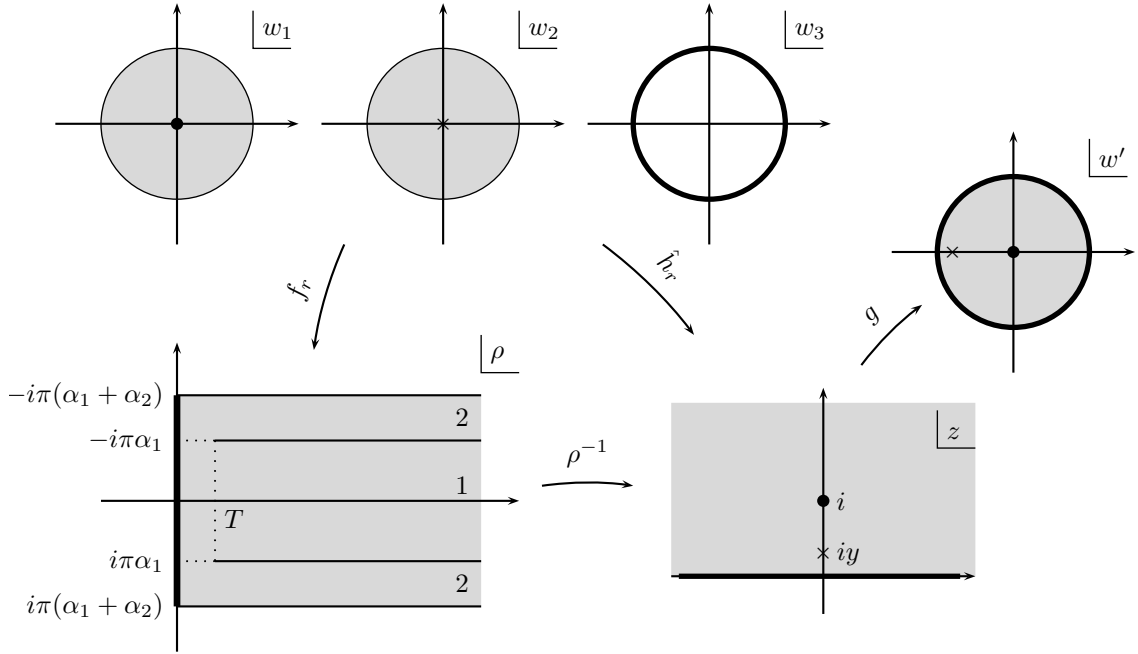}
\caption{Series of conformal mappings from each worldsheet to a unit disk.}
\label{confmaps}
\end{figure}

We construct the $\rho$-plane by gluing the two
unit disks $\abs{w_r}\le1$ $(r=1,2)$ with closed string insertions
and the boundary $\abs{w_3}=1$ together,
\begin{equation}
\rho=f_r(w_r)=\begin{cases}
\al_r\log w_r+T+i\be_r&(r=1,2)\\
\al_3\log w_3+i\be_3&(r=3)
\end{cases},\qquad
(\be_r=\pi\sgn(\arg w_r)\sum_{i=1}^{r-1}\al_i)
\end{equation}
where $T>0$ is a regularization parameter eventually taken to zero.
The $\rho$-plane and the subsequent conformal transformations are depicted
in Figure \ref{confmaps}.
We take the argument of $w_r$ to be from $-\pi$ to $\pi$.
The $\rho$-plane is mapped to the upper-half plane of $z$
by the Mandelstam mapping,
\begin{equation}
\rho(z)=\al_1\log\frac{z-i}{z+i}+\al_2\log\frac{z-iy}{z+iy}\,,\qquad
(0<y<1)
\end{equation}
where $y$ is a parameter describing where the second closed string
insertion is mapped.
In this way, we obtain the LPP vertex in the current case:
\begin{equation}
\LPPT{\Psi(\al_1)}{\Psi(\al_2)}{B(\al_3);\la}
=\Ev{\hat h_1[\Psi(0)]\hat h_2[\Psi(0)]}_\la,\label{PPBT}
\end{equation}
where $\hat h_r(w_r)=\rho^{-1}(f_r(w_r))$ $(r=1,2,3)$ and the subscript $\la$
denotes the existence of
an appropriate conformal transformation of the boundary deformation.
We note that \eqref{PPBT} should coincide with \eqref{PPB} in the
$T\to0$ limit.
The inverses of $\hat h_r(w_r)$ are explicitly given by
\begin{align}
w_1&=\frac{z-i}{z+i}\left(\frac{z-iy}{z+iy}\right)^{\frac{\al_2}{\al_1}}
e^{-\frac{T}{\al_1}}\,,\qquad
w_2=\left(\frac{1+iz}{1-iz}\right)^{\frac{\al_1}{\al_2}}
\left(\frac{z-iy}{z+iy}\right)e^{-\frac{T}{\al_2}}\,,\nt
w_3&=\exp\left[2i\left(\frac{\al_1}{\al_3}\arctan z
+\frac{\al_2}{\al_3}\arctan\frac zy\right)\right].\label{w3}
\end{align}
The interaction point $z_I$ is a solution to $d\rho/dz=0$.
We easily see that the interaction point is on the imaginary axis,
\begin{equation}
z_I=i\sqrt{\frac{(\al_1y+\al_2)y}{\al_1+y\al_2}}\,.
\end{equation}
Then the parameter $y$ is determined via
\begin{equation}
T=\Re\rho(z_I)=\al_1\log\Abs{\frac{z_I-i}{z_I+i}}
+\al_2\log\Abs{\frac{z_I-iy}{z_I+iy}}\,.
\end{equation}
Since both terms are positive, these two terms should vanish in the $T\to0$
limit. This leads to $z_I\to0$ and hence $y\to0$. Moreover, when $y\to0$,
we see from
\eqref{w3} that every point on the upper-half plane has a corresponding point
in the first worldsheet $\abs{w_1}\le1$.
Therefore, in this limit,
the worldsheet of the second string collapses to a point, $z=0$.

Along the argument in \cite{Gaiotto:2002kf}, we find that the correlation
function factors into the product of correlation functions of each
closed string in this limit.
As mentioned above, the surface is splitting into two surfaces when
$T$ approaches zero. The key fact here is that in the splitting limit,
there exists a conformal map which is the identity for one surface
and maps the other surface to a point.
This is fulfilled by further mapping the upper-half plane into a unit disk as
\begin{equation}
w'=g(z)=\frac{z-i}{z+i}=-\exp(2i\arctan z)\,.\label{w'}
\end{equation}
Then the resulting composite map has the desired property,
\begin{equation}
g(\hat h_1(w_1))\to w_1,\qquad
g(\hat h_2(w_2))\to-1.\qquad (T\to0)
\end{equation}
Therefore, in this limit, the conformal mapping leaves the first string
invariant, while the second string is mapped to a point on the boundary,
which can be regarded as an operator insertion on the boundary
of the first worldsheet.
In addition, the projector $\cP$ defined in \eqref{Proj}, in other words,
the level matching condition effectively integrates the operator insertion
over the boundary of $w'$. Thus the boundary insertion must be expanded
in terms of the basis of the integrated boundary operators $O_i$.
If we consider the case where the boundary deformation is
marginal, i.e., the deformation is invariant under the conformal
transformation $g\circ\hat h_3$,
we obtain \eqref{PPBT} in this limit as\footnote{Here
the boundary state is integrated over the length $\al$,
$\bs=\int\frac{d\al}{2\pi}\bra{B(\al)}$. The symmetrization in \eqref{factor1}
and subsequent equations is necessary to reproduce CFT calculations.
\label{lengthBS}}
\begin{equation}
\bssym{O_i}\ket{\Psi(\al_1)}\,F_2^i(\Psi(\al_2))\,,\label{factor1}
\end{equation}
where $F_2^i(\Psi(\al_2))$ is a function of the second closed string field.

We can consider another mapping by exchanging the first and the second
closed strings.
In this case, the first worldsheet is mapped to a point. Then, with a
function $F_1^j(\Psi(\al_1))$, \eqref{PPB} becomes
\begin{equation}
\bssym{O_j}\ket{\Psi(\al_2)}\,F_1^j(\Psi(\al_1))\,.
\end{equation}
Combining the results above, we obtain a factorized form of \eqref{PPB},
\begin{equation}
\bssym{O_i}\ket{\Psi(\al_1)}f^{ij}\bssym{O_j}\ket{\Psi(\al_2)}\,.
\label{factorization}
\end{equation}
This expression should be invariant under reparametrizations of
the space of boundary operators, $\la^i\to \la'^i(\la)$. Thus we require
$f^{ij}$ (with an appropriate sign factor) to be a tensor such that
\eqref{factorization} has the correct ghost number. This uniquely determines
$f^{ij}$ up to an overall constant $C$,
\begin{equation}
f^{ij}=C\om_o^{ij}(-1)^{j+1},
\end{equation}
where $(-1)^{j+1}$ arises when $O_j$ moves across $\bs$.

We can determine the constant $C$ as follows.
If we substitute $c_0^-\ket{\Psi(\al_1)}_1$ and
$\ket{0}_{\al_2}\equiv\ket0_2\otimes\ket{\al_2}_2$ into
$\ket{\Psi(\al_1)}_1$ and $\ket{\Psi(\al_2)}_2$, respectively,\footnote{
To be precise, we must multiply $c_0^-\ket\Psi$ by a spacetime
field of ghost number $-1$ to get the correct total ghost number. In this
sense, we also multiply $\ket0$ by a spacetime ghost field of ghost number 2.}
we can evaluate \eqref{factor1} directly due to
the absence of a closed string insertion on the second worldsheet:
\begin{equation}
\bra\VLPP c_0^{-(1)}\ket{\Psi(\al_1)}_1\ket{0}_{\al_2}\ket{B(\al_3);\la}_3
=\bsla c_0^-\ket{\Psi(\al_1)}\,.\label{fixC1}
\end{equation}
On the other hand, \eqref{factorization} becomes
\begin{align}
C(-1&)^{j+1}\bssym{O_i}c_0^-\ket{\Psi(\al_1)}\om_o^{ij}\bssym{O_j}\ket0\nt
&=C(-1)^{j+1}\bssym{O_i}c_0^-\ket{\Psi(\al_1)}\om_o^{ij}\bssym{O_j,1}\ket0\nt
&=C\bssym{O_i}c_0^-\ket{\Psi(\al_1)}\om_o^{ij}\om^o_{j1}
=C\bsla c_0^-\ket{\Psi(\al_1)}\,,\label{fixC2}
\end{align}
where $O_1\equiv1$. Comparing \eqref{fixC1} with \eqref{fixC2}, we get $C=1$.
We can treat the case with $\al_1,\al_2>0$ and $\al_3<0$ in the same way.
Therefore we obtain, for $\al_1\al_2>0$,
\begin{align}
&\ev{\VLPP|\Psi(\al_1)}_1\ket{\Psi(\al_2)}_2\ket{B(\al_3);\la}_3\nt
&\quad=(-1)^{j+1}
\bssym{O_i}\ket{\Psi(\al_1)}\om_o^{ij}\bssym{O_j}\ket{\Psi(\al_2)}\,,
\end{align}
which leads to a conclusion that \eqref{BV12} is satisfied:
\begin{align}
&\bsla c_0^-\ket{\Psi(\al_1)*\Psi(\al_2)}\nt
&\quad=(-1)^{j+1}\bssym{O_i}\ket{\Psi(\al_1)}\om_o^{ij}
\bssym{O_j}\ket{\Psi(\al_2)}\,, \label{fac2}
\end{align}
where we used \eqref{defstar} and integrated over $\al_3$.

For off-shell boundary deformations, it is in general difficult to
consider the conformal transformation of the deformed boundary state
because the transformation laws of general boundary operators are complicated.
However, in the $T\to0$ limit, the worldsheet just represents two
cylinders\footnote{There is an ambiguity to define amplitudes
when we consider more than one kind of off-shell boundary.
For example, in one-loop BSFT \cite{1loop}, the results depend on the choice
of the Weyl factors in the two boundaries.
In order to avoid the ambiguity consistently, one-loop BSFT is defined to
reproduce standard cylinder amplitudes.
In this sense, when we consider off-shell deformations, we also work on
a cylinder-like coordinate such as the $\rho$-plane.}
attached each other only at a point on the boundary and hence
the closed string fields $\ket{\Psi(\al_1)}_1$ and $\ket{\Psi(\al_2)}_2$
would individually feel the original off-shell boundary
state $\bsla$. It is therefore conceivable that
\eqref{PPB} is given by the two overlaps between
the original boundary state and each closed string field
with an operator inserted on the each boundary, which takes the same form as
\eqref{fac2}.
To make this more concrete, we provide some consistency checks on \eqref{fac2}
in Appendix \ref{consistency}: (i) for the case of constant tachyon,
\eqref{fac2} is reproduced using the result for the undeformed boundary
state.
(ii) The r.h.s. of \eqref{fac2} satisfies the properties of the HIKKO star
product, \eqref{commutative}--\eqref{starJacobi}.
It is crucial that
the off-shell boundary states appeared in the r.h.s. of \eqref{fac2}
are the original one;
otherwise it would break the derivation law
\eqref{derivation} and the Jacobi identity \eqref{starJacobi}.
Therefore it is convincing that \eqref{fac2} holds even for off-shell
boundary states, and hereafter we assume that this is the case.

Finally we comment on the case $\al_1\al_2<0$. This corresponds to the
interchange of ``2'' and ``3'' in Figure \ref{figvertex}.
One can easily see that no factorization occurs since there are no pinching
points on the surface. Hence even for the HIKKO vertex, the relation
\eqref{BV12} does not hold in this case.
One might think that non-polynomial closed SFT \cite{Zwiebach:1992ie,nonpoly2}
is more compatible with BSFT since there are no string length parameters.
The 3-closed-string vertex in
this SFT is a closed string extension of the vertex in cubic open SFT.
Unlike the joining-splitting type vertex as in HIKKO,
the interaction of two closed
strings is not point-like. Thus non-polynomial closed SFT would not
be the closed string part of our open-closed SFT.

\section{Open-closed SFT action and gauge transformation}\label{gauge}
We have seen that, for $\al_1\al_2>0$, the HIKKO 3-string vertex satisfies
\eqref{BV12} and consequently the (second) master equation \eqref{OCBV1}.
Thus we conclude that the closed string sector of our open-closed SFT
is the HIKKO closed SFT,
\begin{equation}
S_c=\frac12\bra{\Psi}c_0^-Q_B\ket{\Psi}
+\frac{\ka^2}3\bra\Psi c_0^-\ket{\Psi*\Psi}\,,\label{actHIKKO}
\end{equation}
where the string field $\ket\Psi$ should be understood as
\begin{equation}
\int_{-\infty}^\infty\frac{d\al}{2\pi}\ket{\Psi(\al)}\,.
\end{equation}
Of course, the closed SFT action \eqref{actHIKKO} with the HIKKO
3-string interaction satisfies the ``closed'' master equation \eqref{OCBV0}.

The remaining issue is the treatment for the contribution from the moduli region with $\al_1\al_2<0$, which does not satisfy the master equation.
One possibility might be adding more terms to the action so that it satisfies the master equation.
In this paper, however, we do not pursue this possibility but just limit our attention within the region where the master equation is satisfied and investigate the implication of the results so far.
As mentioned in footnote \ref{lengthBS}, the boundary state in
\eqref{BV12} should be supplemented with an integral on the length
parameter $\al$.
If we restrict the integration region to $\al>0$,\footnote{$\bra{B(\al)}$
is the BPZ conjugate of $\ket{B(\al)}$, i.e.
$\bra{B(\al)}=\bs\otimes\bra{-\al}$.}
\begin{equation}
\bs\to\Bs\equiv\int_0^\infty\frac{d\al}{2\pi}\bra{B(\al)}\,,
\label{mod_bs}
\end{equation}
we can discard the undesired contribution in the first term of \eqref{BV12} while keeping only the contribution with $\al_1<0$ and $\al_2<0$.
However, this seemingly ad hoc treatment does not solve all the problems
arising from the moduli region.
In the second term of \eqref{BV12}, the modification \eqref{mod_bs}
leaves the contribution with $\al_1+\al_2<0$, which still contains the region with $\al_1\al_2<0$.
As we will see below, this results in a further constraint on
the gauge parameter.
Thus we consider the following open-closed SFT action:
\begin{equation}
S=S_c+\sint+\sbsft,\label{totalact}
\end{equation}
where $S_c$ and $\sbsft$ are already given in \eqref{actHIKKO}
and \eqref{act-BSFT}, respectively, while $\sint$ is \eqref{Sint}
with the replacement \eqref{mod_bs},
\begin{equation}
\sint=\frac12\Bsv c_0^-\ket\Psi\,.
\end{equation}
In the rest of this section, we examine the gauge invariance of this
action.\footnote{Unlike Zwiebach's open-closed SFT \cite{Zwiebach:1997fe},
open-closed interaction terms of higher order in the closed string field
would break the closed string gauge invariance
since the closed string part is cubic.}

Remember that the open-closed master equation leads to the constraint
\eqref{constraint} on the gauge parameter, even if the master equation is
fully satisfied.
Let us discuss this constraint first.
Since $\sint$ is linear in closed string fields
and the two-form $\om_c$ is constant, the constraint reduces to
\begin{align}
0&=\frac{\p_r}{\p\la^k}\left(\frac{\p_l\sint}{\p\psi^J}\right)\ep^k
=\frac{(-1)^J}2\frac{\p_r}{\p\la^k}\Bigl(\Bsv c_0^-\ket{\Psi_J}\Bigr)\ep^k\nt
&=-i\ka\Bssym{[b_0^-,\La_o]}c_0^-\ket{\Psi_J}\,,
\end{align}
where $\La_o\equiv O_k\ep^k$ is the gauge parameter
for open string fields. This constraint is easily solved by requiring
\begin{equation}
[b_0^-,\La_o]=0\,,\label{b0Lambda}
\end{equation}
without imposing any constraints on $\ep^K$. Once this constraint is imposed,
it follows that
\begin{equation}
\frac{\p_r\om^o_{ij}}{\p\la^k}\ep^k
=\frac{\p_r\om_o^{ij}}{\p\la^k}\ep^k
=\frac{\p_r\sint}{\p\la^k}\ep^k=0\,,
\end{equation}
which simplifies \eqref{open} as
\begin{align}
\de_o\la^i&=\om_o^{ij}\left(\frac{\p_r}{\p\la^k}
\frac{\p_l\sbsft}{\p\la^j}\right)\ep^k
=\om_o^{ij}\frac{\p_r(\om^o_{jl}V^l)}{\p\la^k}\ep^k
=\frac{\p_rV^i}{\p\la^k}\ep^k\,.
\end{align}
Then the open string gauge transformation is given by
\begin{align}
\delo O&=O_i\,\de_o\la^i=O_i\frac{\p_rV^i}{\p\la^k}\ep^k
=\frac{\p_r(O_iV^i)}{\p\la^k}\ep^k
=\frac{\p_r[Q_B,O]}{\p\la^k}\ep^k
=[Q_B,\La_o]\,.\label{deloO}
\end{align}
Despite the fact that the action \eqref{totalact} does not completely satisfy
the open-closed master equation, one can easily check that the action is
indeed invariant under the gauge transformation as follows.
We find that the off-shell boundary state is invariant,
\begin{equation}
\delo\bsv=2i\ka\bssym{[b_0^-,[Q_B,\La_o]]}=0\,,
\end{equation}
where we used the Jacobi identity \eqref{Jac}, the constraint condition
\eqref{b0Lambda}, as well as the fact that the gauge parameter $\La_o$ is
expanded in terms of the basis $O_i$ satisfying $[L_0^-,O_i]=0$.
Obviously, closed string fields do not transform under the gauge transformation
by $\La_o$.
Thus it is straightforward to show that the action is invariant under
the open string gauge transformation:
\begin{align}
&\delo S_c=0,\qquad\delo\sint=0\,,\\
&\delo\sbsft=\frac{i}{2\ka}\bssym{[Q_B,[Q_B,\La_o]]}c_0^-\ket0=0\,.
\end{align}

Let us turn to the closed string gauge transformation. In this case,
the constraint \eqref{constraint} keeps the gauge transformation intact
and \eqref{closed} gives the following gauge transformation law:
\begin{align}
\delc\ket{\Psi}&=Q_B\ket{\La_c}+2\ka^2\ket{\Psi*\La_c}\,,\label{gaugetrclosed}\\
\delc O
&=i\ka(-1)^{j+1}O_i\,\om_o^{ij}\Bssym{O_j}\ket{\La_c}\,.\label{delcO}
\end{align}
Here $\ket{\La_c}\equiv-\ket{\Psi_K}\ep^K$ is the parameter
for the closed string gauge transformation, which is of ghost number 1
and annihilated by both $b_0^-$ and $L_0^-$.
In the following, we decompose the
gauge parameter $\ket{\La_c}$ as
\begin{equation}
\ket{\La_c}=\ket{\La_c^-}+\ket{\La_c^+}\,,
\end{equation}
where
\begin{equation}
\ket{\La_c^-}=\int_{-\infty}^0\frac{d\al}{2\pi}\,\ket{\La_c(\al)}\,,\qquad
\ket{\La_c^+}=\int_0^\infty\frac{d\al}{2\pi}\,\ket{\La_c(\al)}\,.
\end{equation}
Similarly, we decompose $\ket\Psi=\ket{\Psi^-}+\ket{\Psi^+}$.

First, let us evaluate the effect of the transformation of open string fields under the closed string gauge transformation.
The gauge transformation of the BSFT action $\sbsft$ is given by
\begin{align}
\delc\sbsft
&=-\bssym{[Q_B,O],\delc O}\ket0\nt
&=-i\ka(-1)^{j+1}\bssym{[Q_B,O],O_i}\ket0\,\om_o^{ij}\,
\Bssym{O_j}\ket{\La_c}\nt
&=-i\ka\Bssym{[Q_B,O]}\ket{\La_c}\nt
&=\frac12\Bsla Q_Bc_0^-\ket{\La_c^-}\,.\label{delcSBSFT}
\end{align}
See \eqref{complete2} for the calculation details including $\om_o^{ij}$.
On the other hand, the open-closed interaction term $\sint$ is
transformed by the gauge transformation of open string fields as
\begin{align}
\frac{1}{2}\big(\delc&\Bsv\big)c_0^-\ket\Psi
=i\ka\Bssym{\delc O}\ket{\Psi^-}\nt
&=-\ka^2(-1)^{j+1}\Bssym{O_i}\ket{\Psi^-}\om_o^{ij}\Bssym{O_j}\ket{\La_c}\nt
&=-\ka^2(-1)^{j+1}\Bssym{O_i}\ket{\Psi^-}\om_o^{ij}\Bssym{O_j}\ket{\La_c^-}\nt
&=-\ka^2\Bsla c_0^-\ket{\Psi^-*\La_c^-}\,.\label{delSintbyopen}
\end{align}
Here, the last equality comes from the following relation,
\begin{equation}
(-1)^{j+1}\Bssym{O_i}\ket{\Psi^-}\om_o^{ij}
\Bssym{O_j}\ket{\La_c^-}=\Bsla c_0^-\ket{\Psi^-*\La_c^-}\,, \label{modBV12}
\end{equation}
which is obtained by applying $\overleftarrow{\p_K}\ep^K$ on the both
sides of \eqref{BV12} with $\alpha_1<0$ and $\alpha_2<0$.
Note that the transformation \eqref{delSintbyopen} implies that the gauge transformation for the boundary state is given by\footnote{
This gauge transformation law for boundary states
was first obtained in \cite{Hashimoto:1997vz}
for on-shell boundary deformations.}
\begin{equation}
\delc\Bsla=2\ka^2\Bsla*\bra{\La_c^-}\,,\label{delcmBS}
\end{equation}
which can be obtained by using the cyclic property \eqref{cyclic}
as well as \eqref{inner}.

Next we consider the effect of the gauge
transformation of closed string fields. It is obvious that $S_c$ is invariant,
whereas the contribution from $\sint$ gives
\begin{align}
\frac12&\Bsla c_0^-\delc\ket\Psi\nt
&=\frac12\Bsla c_0^-Q_B\ket{\La_c}
+\ka^2\Bsla c_0^-\ket{\Psi^-*\La_c}\nt
&=\frac12\Bsla c_0^-Q_B\ket{\La_c^-}
+\ka^2\Bsla c_0^-\ket{\Psi^-*\La_c^-}+\ka^2\Bsla c_0^-\ket{\Psi^-*\La_c^+}\,.
\label{delSint}
\end{align}
The first two terms exactly cancel the previous contributions, \eqref{delcSBSFT} and \eqref{delSintbyopen}, from the gauge transformation of open string fields.\footnote{Note that $\Bsla c_0^-Q_B\ket{\La_c^-}
=\Bsla c_0^-Q_Bb_0^-c_0^-\ket{\La_c^-}=-\Bsla Q_Bc_0^-\ket{\La_c^-}$.}
However, unfortunately, the last term is not canceled by any other term.
This is because our open-closed SFT action does not completely satisfy
the master equation.
Thus the action \eqref{totalact} is invariant only if we restrict the closed
string gauge transformation parameter $\ket{\La_c}$ to $\ket{\La_c^+}=0$.

In summary, we have shown that the action \eqref{totalact} is invariant
under the following gauge transformations; $\delo S=\delcm S=0$ with
\begin{align}
&\delo O=[Q_B,\La_o]\,,\qquad(\text{with }[b_0^-,\La_o]=0)\\
&\delcm\ket{\Psi}=Q_B\ket{\La_c^-}+2\ka^2\ket{\Psi*\La_c^-}\,,\label{delcmPsi}\\
&\delcm O=i\ka(-1)^{j+1}O_i\,\om_o^{ij}\Bssym{O_j}\ket{\La_c^-}\,.
\label{delcmO}
\end{align}
It is worthwhile to note that, though \eqref{delcmO}
looks unfamiliar, the resultant gauge transformation for the off-shell boundary state
\eqref{delcmBS} takes the same form as the nonlinear part of \eqref{delcmPsi}.

\section{Conclusion and Discussion}
We have investigated the open-closed SFT whose open string sector is BSFT.
In order to determine the closed and open-closed string interactions,
we used the BV classical open-closed master equations.
Once we introduce the conventional open-closed interaction term $\sint$,
the master equations determine the closed string sector to be the HIKKO
closed SFT. We have also derived the gauge transformation law for both
open and closed string fields.
In solving the master equation, we have assumed that \eqref{fac2} holds
even for off-shell boundary states. Though we have provided several
evidences to support this, it is of course important to show the relation
directly. However this is technically difficult and we leave it for future work.

In our open-closed SFT, the gauge parameters must be supplemented
with the constraints $[b_0^-,\La_o]=0$ and $\ket{\La_c^+}=0$.
While the former is simply due to the classical truncation of the full
master equation, the latter constraint comes from
the existence of the parameter region where
the classical open-closed master equation is not satisfied.
This is not satisfactory since the gauge invariance
of the original closed SFT is half broken. One might conclude that there is no
consistent way to couple BSFT to closed SFT with the conventional open-closed
interaction term $\sint$. However,
we can find a hint from a similar case in
\cite{Baba:2006rs,Baba:2007tc}, where
D-branes are realized as soliton states
constructed on the second quantized vacuum $\dket0$
in the $OSp$ invariant closed SFT.
The D-brane soliton state roughly takes the form
\begin{equation}
\exp\left[-\sint+\cdots\right]\dket0\,.\label{solitonic}
\end{equation}
Here, the vacuum $\dket0$ is defined to be
annihilated by string fields with positive length, $\ket{\Psi(\al)}\dket0=0$
for $\al>0$.
Since the 3-string vertex in this closed SFT is the same as
the HIKKO vertex, it might be possible to interpret our results in the
context of this D-brane soliton state.
Then the integration region of $\bra{B(\al)}$ in the $\sint$ is
naturally restricted to $\alpha>0$ as in \eqref{mod_bs}.
Moreover, the vacuum $\dket0$ annihilates $\ket{\La_c^+}$,
which would make the constraint $\ket{\La_c^+}=0$ automatic
and hence unnecessary.\footnote{
In order to show the gauge invariance of the D-brane soliton, it would be
necessary to use the idempotency relation for boundary states
\cite{Kishimoto:2003ru,Kishimoto:2003yu,Kishimoto:2004km} as in
\cite{Baba:2007tc}.}
In this scenario, the BSFT action would arise through
the off-shell extension of the D-brane soliton state.
In other words, $\sbsft$ may appear at ``$\cdots$'' in
\eqref{solitonic}.
If this is the case, BSFT might be considered as the effective theory
describing the fluctuations of the D-brane soliton state.

Finally, it is worth mentioning the relation of the present work to
\cite{Hashimoto:1997vz}.
The authors considered the HIKKO closed SFT with an
on-shell D-brane source term and found that the gauge invariance requires
the DBI action. They showed that
the total action correctly reproduces the low-energy dynamics of a D-brane.
Remarkably, our SFT action takes the same form of the action in
the reference with the replacement of the DBI action by BSFT
as well as with the off-shell extension of the source term.\footnote{
The open-closed interaction term here is linear in the closed string field.
One might wonder how such a linear term could reproduce the non-linear coupling
between the graviton and matter. Actually, a similar question arises
in the original HIKKO closed SFT, because it only contains
cubic interaction terms while the Einstein-Hilbert action involves
all orders in the metric.
A plausible scenario would be that,
in these SFTs, the component fields are related to fields
in the corresponding effective field theories through non-trivial field
redefinition as well as integration of massive fields. In
\cite{Hashimoto:1997vz}, the authors examined such a correspondence at the
first non-trivial order.}
Thus our open-closed SFT is a natural off-shell extension of their work
and we expect that it describes the off-shell dynamics
of string theory such as D-brane decay by tachyon condensation
and closed string radiation.
We hope that our formulation will provide new insights into nonperturbative
aspects of string theory.

\section*{Acknowledgments}
We would like to thank Hiroyuki Hata and Yuji Okawa for useful discussions
at the early stage of this work.
\section*{Appendix}
\appendix
\section{Conventions and useful formulas}\label{convention}
In this paper we mainly follow the conventions in \cite{Polchinski}
and \cite{Zwiebach:1992ie}. We use the unit $\al'=2$.\\
Virasoro generators:
\begin{equation}
L_m=L_m^\text m+L_m^\text g\,,
\end{equation}
where
\begin{equation}
L_m^\text m=\frac12\sum_n\normalorder{\al^\mu_{m-n}\al_{\mu n}}\,,\qquad
L_m^\text g=\sum_n(2m-n)\normalorder{b_nc_{m-n}}-\de_{m,0}\,.
\end{equation}
BRST operator:
\begin{equation}
Q_B=\sum_n c_{-n}L_n^\m+\sum_{m,n}\frac{m-n}2\normalorder{b_{-n-m}c_mc_n}
-c_0+\text{(anti-holomorphic part)}\,.
\end{equation}
Ghost zero-modes:
\begin{align}
b_0^+\equiv&b_0+\tilde b_0\,,&b_0^-\equiv&\frac12(b_0-\tilde b_0)\,,&
c_0^+\equiv&\frac12(c_0+\tilde c_0)\,,&c_0^-\equiv&c_0-\tilde c_0\,.
\end{align}
Normalization:
\begin{equation}
\bra0c_{-1}\tilde c_{-1}c_0^-c_0^+c_1\tilde c_1\ket0=1\,.
\end{equation}
Boundary state for D$p$-brane (located at $x^i=0$):
\begin{equation}
\bs=-\bra0c_{-1}\tilde c_{-1}c_0^+\de^{25-p}(\hat x^i)
\exp\left[-\sum_{n=1}^\infty \frac1n\left(\al_{n}^\mu\ti\al_{\mu n}
-\al_n^i\ti\al_{in}\right)
-\sum_{n=1}^\infty(c_{n}\ti b_{n}+\ti c_n b_n)\right],
\end{equation}
where $\mu=0,\ldots,p$ and $i=p+1,\ldots,25$. This satisfies
the following conditions:
\begin{equation}
\bs(\al_n^\mu+\ti\al_{-n}^\mu)
=\bs(\al_n^i-\ti\al_{-n}^i)
=\bs(c_n+\ti c_{-n})=\bs(b_n-\ti b_{-n})=0\,.
\end{equation}
Jacobi identity:
\begin{equation}
(-1)^{AC}\big[A,[B,C]\big]+\text{cyclic perms}=0\,.\label{Jac}
\end{equation}
Inner products of string fields:
\begin{equation}
\bra Ac_0^-\ket B=(-1)^{(A+1)(B+1)}\bra Bc_0^-\ket A,\qquad
\bra Ac_0^-Q_B\ket B=(-1)^{AB}\bra Bc_0^-Q_B\ket A\,.\label{inner}
\end{equation}
Derivatives of $\sfr$ and $\sint$:
\begin{align}
\frac{\p_r\sfr}{\p\psi^I}&=(-1)^I\frac{\p_l\sfr}{\p\psi^I}
=\bra\Psi c_0^-Q_B\ket{\Psi_I}=\bra{\Psi_I}c_0^-Q_B\ket\Psi\,,\label{delSfree}\\
\frac{\p_r\sint}{\p\la^i}&=(-1)^i\frac{\p_l\sint}{\p\la^i}
=i\ka\bssym{O_i}\ket\Psi\,.
\end{align}
Component form of \eqref{defofS}:
\begin{equation}
\frac{\p_l\sbsft}{\p\la^i}=(-1)^i\frac{\p_r\sbsft}{\p\la^i}
=(-1)^{i+1}\bssym{O_i,[Q_B,O]}\ket0
=\om^o_{ij}V^j\,. \label{delSBSFT}
\end{equation}
Calculation in \eqref{delcSBSFT}:
\begin{align}
(-1&)^{j+1}\bssym{[Q_B,O],O_i}\ket0\,\om_o^{ij}\,\Bssym{O_j}\ket{\La_c^-}\nt
&=(-1)^{j+1}\bssym{O_kV^k,O_i}\ket0\,\om_o^{ij}\,\Bssym{O_j}\ket{\La_c^-}\nt
&=(-1)^{j+1}V^k\bssym{O_k,O_i}\ket0\,\om_o^{ij}\,\Bssym{O_j}\ket{\La_c^-}\nt
&=(-1)^{j+k}V^k\om^o_{ki}\,\om_o^{ij}\,\Bssym{O_j}\ket{\La_c^-}\nt
&=V^j\Bssym{O_j}\ket{\La_c^-}\nt
&=\Bssym{[Q_B,O]}\ket{\La_c^-}\,.\label{complete2}
\end{align}
Properties of the anti-bracket:
\begin{align}
&\abs{\{S_1,S_2\}}=\abs{S_1}+\abs{S_2}+1\,,\\
&\{S_1,S_2\}=-(-1)^{(S_1+1)(S_2+1)}\{S_2,S_1\}\,,\label{SS}\\
&(-1)^{(S_1+1)(S_3+1)}\{\{S_1,S_2\},S_3\}+\text{cyclic perms}=0\,.
\label{JacobiAB}
\end{align}
\textbf{Completeness condition for closed string fields}:
\begin{equation}
\ket{\Psi_I}\om_c^{IJ}(-1)^{J+1}\bra{\Psi_J}=b_0^-\cP\,,\label{complete}
\end{equation}
where $\cP$ is the projector for the level matching condition given in
\eqref{Proj}.\\
\textit{Proof}:\quad Any state $\ket\Phi$ in the closed string Hilbert space
can be uniquely decomposed as
\begin{equation}
\ket\Phi=c_0^-\ket{\Psi}+\ket{\Psi_2}+\ket\chi\,,\label{anystate}
\end{equation}
where $\ket{\Psi}$ and $\ket{\Psi_2}$ are states annihilated by
both $b_0^-$ and $L_0^-$, while $\ket\chi$ is annihilated by $\cP$.
Since $\bra{\Psi_J}=\bra{\Psi_J}c_0^-b_0^-\cP$, by
applying the l.h.s. of \eqref{complete} on both sides of \eqref{anystate},
we have
\begin{align}
(-1)^{J+1}\ket{\Psi_I}\om_c^{IJ}\bracket{\Psi_J}{\Phi}
&=(-1)^{J+1}\ket{\Psi_I}\om_c^{IJ}\bra{\Psi_J}c_0^-\ket{\Psi}
=(-1)^{J+1}\ket{\Psi_I}\om_c^{IJ}\bra{\Psi_J}c_0^-\ket{\Psi_K}\psi^K\nt
&=\ket{\Psi_I}\om_c^{IJ}\om^c_{JK}\psi^K
=\ket{\Psi_I}\psi^I
=\ket{\Psi}\,.
\end{align}

\section{HIKKO closed string field theory}\label{HIKKO}
We summarize the basic properties of the HIKKO closed SFT \cite{Hata:1986kj}
in CFT language.
(See also \cite{Kishimoto:2003yu}.)
Unlike the usual convention, we adopt a convention where string fields
have ghost number 2.
The explicit forms of the mappings $h_r$ in \eqref{defLPP} are given by
\begin{equation}
h_r(w_r)=\rho^{-1}(f_r(w_r))\,.\qquad(r=1,2,3)
\end{equation}
Here $f_r$'s glue three unit disks (with coordinates $w_r$) together
to construct the $\rho$-plane,
\begin{equation}
f_r(w_r)=\al_r\log w_r+\tau_0+i\be_r\,,\qquad
\big(\be_r=\pi\sgn(\arg w_r)\sum_{i=1}^{r-1}\al_i\big)
\end{equation}
and $\rho^{-1}$ is the inverse of the Mandelstam mapping
\begin{equation}
\rho(z)=\al_1\log(z-z_1)+\al_2\log(z-z_2)+\al_3\log(z-z_3)\,,
\end{equation}
where we can choose $z_1$, $z_2$ and $z_3$ arbitrarily.
The interaction point $\tau_0$ is given by $\tau_0=\Re\rho(z_0)$, where
$z_0$ is the solution to $d\rho/dz=0$. Using the conformal mappings above,
we define the star product via \eqref{defstar} and \eqref{defLPP}.

The action of the HIKKO closed SFT is given by
\begin{equation}
S=\frac12\Psi\cdot Q_B\Psi+\frac{\ka^2}3\Psi\cdot(\Psi*\Psi)\,,
\end{equation}
where the dot product is defined by
\begin{equation}
\Psi_1\cdot\Psi_2\equiv\bra{\Psi_1}c_0^-\ket{\Psi_2}\,.
\end{equation}
The basic properties of the dot product and the star product are summarized
as follows.
Note that some sign factors are different from the conventional ones.
\begin{align}
&\Psi_1\cdot\Psi_2=(-1)^{(\abs{1}+1)(\abs{2}+1)}\Psi_2\cdot\Psi_1\,,\\
&(Q_B\Psi_1)\cdot\Psi_2=(-1)^{\abs{1}}\Psi_2\cdot(Q_B\Psi_1)\,,\\
&\Psi_1*\Psi_2=(-1)^{\abs1\abs2}\Psi_2*\Psi_1\,,\label{commutative}\\
&Q_B(\Psi_1*\Psi_2)+(Q_B\Psi_1)*\Psi_2+(-1)^{\abs1}\Psi_1*(Q_B\Psi_2)=0\,,
\label{derivation}\\
&(-1)^{\abs1\abs3}(\Psi_1*\Psi_2)*\Psi_3
+(-1)^{\abs1\abs2}(\Psi_2*\Psi_3)*\Psi_1
+(-1)^{\abs2\abs3}(\Psi_3*\Psi_1)*\Psi_2=0\,,\label{starJacobi}\\
&\Psi_1\cdot(\Psi_2*\Psi_3)=(-1)^{(\abs1+\abs2)\abs3}\Psi_3\cdot(\Psi_1*\Psi_2)
=(-1)^{(\abs2+\abs3)\abs1}\Psi_2\cdot(\Psi_3*\Psi_1)\,,
\end{align}
where $\abs i$ denotes the Grassmann parity of $\Psi_i$.

\section{On the relation \eqref{fac2}}\label{consistency}
We demonstrate the consistency of the relation \eqref{fac2}
for off-shell boundary states. For simplicity, we put $\ka=1$.
We first consider the off-shell boundary state in the presence of
a constant tachyon $T$; $\bra{B;\la_t}\equiv\bs e^{-T}$.
We can factor out $e^{-T}$ from the off-shell boundary state as
\begin{equation}
\bra{B;\la_t} c_0^-\ket{\Psi(\al_1)*\Psi(\al_2)}
=e^{-T}\bs c_0^-\ket{\Psi(\al_1)*\Psi(\al_2)}\,,
\end{equation}
which can be evaluated by using the result for the undeformed boundary state:
\begin{align}
&e^{-T}(-1)^{j+1}\bs O_i\ket{\Psi(\al_1)}\,\om_o^{ij}\big|_{\la=0}
\bs O_j\ket{\Psi(\al_2)}\nt
&\quad=(-1)^{j+1}\bs\Sym\bigl[e^{-T};O_i\bigr]\ket{\Psi(\al_1)}\,\om_o^{ij}
\bs\Sym\bigl[e^{-T};O_j\bigr]\ket{\Psi(\al_2)}\,,
\end{align}
where note that $\om_o^{ij}$ is proportional to $e^{T}$.
Therefore, though $\bra{B;\la_t}$ is not invariant under the conformal
transformation, we found that \eqref{fac2} is correctly reproduced.

We next show that \eqref{fac2} is consistent with the properties
of the HIKKO star product for any off-shell boundary state.
Hereafter we use an abbreviated notation,
\begin{equation}
S_i\equiv\sint\Big|_{\Psi=\Psi_i(\al_i)}
=\frac12\bsla c_0^-\ket{\Psi_i(\al_i)}\,.\qquad(i=1,2,\dots)
\end{equation}
where $\al_i$'s are taken to be negative.
We now demonstrate that the properties of the HIKKO vertex
\eqref{commutative}--\eqref{starJacobi} are reproduced, even if we use
the r.h.s. of the following relation,
\renewcommand{\ka}{}
\begin{align}
\bsla&c_0^-\ket{\Psi_1(\al_1)*\Psi_2(\al_2)}\nt
&=(-1)^{(i+1)\abs1+j+1}\bssym{O_i}\ket{\Psi_1(\al_1)}\om_o^{ij}
\bssym{O_j}\ket{\Psi_2(\al_2)}\nt
&=(-1)^{\abs1}\{S_1,S_2\}_o\,,\label{BPP}
\end{align}
instead of the original HIKKO vertex.
Here $(-1)^{\abs1}$ is the Grassmann parity of $\ket{\Psi_1(\al_1)}$,
which arises when $\ket{\Psi_1(\al_1)}$ moves from next to
$\ket{\Psi_2(\al_2)}$ to the position in the second line.
Though closed string fields
are Grassmann even, this expression is necessary for later analysis.

The commutativity \eqref{commutative} is obvious from \eqref{SS}:
\begin{equation}
\bsla c_0^-\ket{\Psi_1*\Psi_2}=\{S_1,S_2\}_o
=\{S_2,S_1\}_o=\bsla c_0^-\ket{\Psi_2*\Psi_1}\,,
\end{equation}
where $\ket{\Psi_i}\equiv\ket{\Psi_i(\al_i)}$.
The derivation law \eqref{derivation} is obtained as follows.
Using \eqref{compV}, \eqref{delSBSFT} and the Jacobi identity
\eqref{JacobiAB} as well, we have
\begin{align}
\bsla c_0^-Q_B\ket{\Psi_1*\Psi_2}
&=-\frac{\p_r}{\p\la^i}\big(\bsla c_0^-\ket{\Psi_1*\Psi_2}\big)\,V^i\nt
&=-\frac{\p_r}{\p\la^i}\big(\{S_1,S_2\}_o\big)\,\om_o^{ij}\om^o_{jk}V^k\nt
&=-\{\{S_1,S_2\}_o,\sbsft\}_o\nt
&=\{\{S_1,\sbsft\}_o,S_2\}_o+\{\{S_2,\sbsft\}_o,S_1\}_o\,.
\end{align}
The two terms are rewritten using
\begin{equation}
\{S_i,\sbsft\}_o=\frac12\bsla c_0^-Q_B\ket{\Psi_i}\,,
\end{equation}
and then we find
\begin{equation}\label{del_C}
\bsla c_0^-Q_B\ket{\Psi_1*\Psi_2}
+\bsla c_0^-\ket{Q_B\Psi_1*\Psi_2}+\bsla c_0^-\ket{\Psi_1* Q_B\Psi_2}=0\,.
\end{equation}
In addition, as
\begin{equation}
\bsla c_0^-\ket{(\Psi_1*\Psi_2)*\Psi_3}=-\{\{S_1,S_2\}_o,S_3\}_o\,,
\end{equation}
the Jacobi identity \eqref{JacobiAB} again leads to
\begin{equation}
\bsla c_0^-\big[\,\ket{(\Psi_1*\Psi_2)*\Psi_3}
+\ket{(\Psi_2*\Psi_3)*\Psi_1}+\ket{(\Psi_3*\Psi_1)*\Psi_2}\big]=0\,,
\label{Jac_C}
\end{equation}
which is consistent with \eqref{starJacobi}.
Thus we have correctly recovered the properties of the HIKKO star product
\eqref{commutative}--\eqref{starJacobi} without referring to
the original HIKKO vertex.
It is important to note that, in both \eqref{del_C} and \eqref{Jac_C},
the equalities are guaranteed by the Jacobi identity \eqref{JacobiAB}
among $\sint$ and $\sbsft$.
In other words, the form of the r.h.s. of \eqref{BPP} is crucial to
reproduce the properties of the HIKKO vertex.

\end{document}